\documentclass[useAMS,usenatbib]{mn2e}
\usepackage{epsfig}
\usepackage{amsmath}
\usepackage{multirow}
\usepackage[english]{babel}
\usepackage{url}
\usepackage{amssymb}
\usepackage{graphicx}
\usepackage{graphics}
\usepackage{natbib}
\usepackage{booktabs}

\newcommand{\OII}{[\mbox{O\,{\sc ii}}]}
\newcommand{\OIIIb}{[\mbox{O\,{\sc iii}}]}
\newcommand{\NeIII}{[\mbox{Ne\,{\sc iii}}]}
\newcommand{\NII}{[\mbox{N\,{\sc ii}}]}

\newcommand{\Msun}{\ensuremath{M_{\odot}}}
\newcommand{\Ha}{\ensuremath{H\alpha}}

\title[GRB host galaxies with VLT/X-Shooter]{GRB host galaxies with VLT/X-Shooter: properties at $0.8 < \mathrm{z} <1.3$\thanks{Based on observations collected at the European Southern Observatory Very Large Telescope, Cerro Paranal, Chile. Programs 084.A-0631(A), 085A-0795(A) and 0.87A-0451(A)}}
\author[S. Piranomonte et al.]{S. Piranomonte$^{1}$ \thanks{E-mail: silvia.piranomonte@oa-roma.inaf.it}, J. Japelj$^{2}$, S. D. Vergani$^{3,4}$, S. Savaglio$^{5,6,7}$, E. Palazzi$^{8}$, S. Covino$^{4}$,\and
H. Flores$^{3}$, P. Goldoni$^{9}$, G. Cupani$^{10}$, T. Kr\"{u}hler$^{11,12}$, F. Mannucci$^{13}$, F. Onori$^{14}$,\and A. Rossi$^{8}$, V. D'Elia$^{1,15}$, E. Pian$^{8,16}$, P. D'Avanzo$^{4}$, A. Gomboc$^{2}$, \and F. Hammer$^{3}$, S. Randich$^{13}$, F. Fiore$^{1}$, L. Stella$^{1}$, G. Tagliaferri$^{4}$\\
$^{1}$ INAF - Osservatorio Astronomico di Roma, via Frascati 33, Monte Porzio Catone (RM), 00040 Italy\\
$^{2}$ Faculty of Mathematics and Physics, University of Ljubljana, Jadranska ulica 19, SI-1000 Ljubljana, Slovenia\\
$^{3}$ Laboratoire Galaxies Etoiles Physique et Instrumentation, Observatoire de Paris, 5 place Jules Janssen, 92195 Meudon, France\\
$^{4}$ INAF - Osservatorio Astronomico di Brera, via E. Bianchi 46, 23807 Merate (LC), Italy\\
$^{5}$ Physics Department, University of Calabria, via P. Bucci, 87036 Arcavacata di Rende, Italy\\
$^{6}$ European Southern Observatory, Karl-Schwarzschild-Strasse 2, 85748 Garching bei Munchen, Germany\\
$^{7}$ Max Planck Institute for Extraterrestrial Physics, Garching, Germany\\
$^{8}$ INAF - Istituto di Astrofisica Spaziale e Fisica Cosmica, Sezione di Bologna, via Gobetti 101, 40129 Bologna, Italy\\
$^{9}$ APC, Univ. Paris Diderot, CNRS/IN2P3, CEA/Irfu, Obs. de Paris, Sorbonne Paris Cité, France\\
$^{10}$ INAF - Osservatorio Astronomico di Trieste, via G.B. Tiepolo 11, 34143 Trieste, Italy\\
$^{11}$ European Southern Observatory, Alonso de Cordova 3107, Vitacura, Casilla 19001, Santiago 19, Chile\\
$^{12}$ Dark Cosmology Centre, Niels Bohr Institute, University of Copenhagen, Juliane Maries Vej 30, 2100 Kobenhavn 0, Denmark\\
$^{13}$ INAF - Osservatorio Astrofisico di Arcetri, INAF, Largo E. Fermi 5, 50125 Firenze, Italy\\
$^{14}$ Dipartimento di Matematica e Fisica, Universit\'a degli Studi Roma Tre, Via della Vasca Navale 84, I-00146 Roma, Italy\\
$^{15}$ ASI-Science Data Center, Via del Politecnico snc, I-00133 Rome, Italy\\
$^{16}$ Scuola Normale Superiore, Piazza dei Cavalieri 7, 56126 Pisa, Italy\\
}

\begin{document}
\date{}

\pagerange{\pageref{firstpage}--\pageref{lastpage}} \pubyear{2015}

\maketitle

\label{firstpage}

\begin{abstract}
Long gamma-ray bursts (LGRBs) are associated with the death of massive stars. Their host galaxies therefore represent a unique class of objects tracing star formation across the observable Universe. Indeed, recently accumulated evidence shows that GRB hosts do not differ substantially from general population of galaxies at high ($z > 2$) redshifts. However, it has been long recognised that the properties of $z < 1.5$ hosts, compared to general star-forming population, are unusual. To better understand the reasons for the supposed difference in LGRB hosts properties at $z < 1.5$, we obtained VLT/X-Shooter spectra of six hosts lying in the redshift range of $0.8 < z < 1.3$. Some of these hosts have been observed before, yet we still lack well constrained information on their characteristics such as metallicity, dust extinction and star formation rate. We search for emission lines in the VLT/X-Shooter spectra of the hosts and measure their fluxes. We perform a detailed analysis, estimating host average extinction, star-formation rates, metallicities and electron densities where possible. Measured quantities of our hosts are compared to a larger sample of previously observed GRB hosts at $z < 2$. Star-formation rates and metallicities are measured for all the hosts analyzed in this paper and metallicities are well determined for 4 hosts. The mass-metallicity relation, the fundamental metallicity relation and SFRs derived from our hosts occupy similar parameter space as other host galaxies investigated  so-far at the same redshift. We therefore conclude that GRB hosts in our sample support the found discrepancy between the properties of low-redshift GRB hosts and the general population of star-forming galaxies.
\end{abstract}


\section{Introduction}

Long duration gamma-ray bursts (LGRBs) originate from the deaths of very massive stars. This is supported both from observational \citep[][]{Galama1998,Hjorth2012} and theoretical \citep{paczynski98, MacFadyen1999, Woosley2006} evidence. Thanks to their exceptional brightness they can be used as powerful sources to unveil the properties of their host galaxies at different epochs in the Universe. LGRB host galaxies form a unique sample of galaxies, covering a wide redshift range. First studies of LGRB hosts found them to be sub-luminous and blue \citep{LeFloch2003}, moderately star-forming (SFR $\sim 1-10 ~{\rm M}_{\odot} ~{\rm yr}^{-1}$), with stellar masses of $\sim 1-5\times10^9 ~{\rm M}_{\odot}$ and a high specific star formation rate (sSFR = SFR/M$_*$) \citep{savaglio09}. Later it was found that the hosts of the so-called dark GRBs \citep{Jakobsson2004,Horst2009} are more dusty and massive \citep{Perley2009,Perley2013,Hunt2014}.  GRB hosts are not selected according to their luminosity, but rather to the fact that a peculiar type of explosion of a massive star has occurred. Therefore, to first order, they trace regions of star formation at all epochs \citep{Kistler2013,Hunt2014}.

LGRBs are associated with broad-lined type Ibc supernovae (SNe) (e.g. \citealt{Hjorth2012} and references therein; see however GRB 060614: \citealt{dellavalle06, fynbo06, galyam06}). It is still not clear which are the conditions that lead a massive star to have the special kind of core collapse that triggers the formation of a jet and a LGRB. Metallicity (which in single star progenitors is directly coupled with rotation) is one of the fundamental parameters predicted to impact the evolution of massive stars as well as their explosive deaths and is expected to play a fundamental role in the formation of LGRBs (e.g. \citealt{Woosley1993, MacFadyen1999, heger03, yoon2010}). Since theoretical models predict LGRBs occur preferentially in a low metallicity environment, a special attention has been given to the study of chemical abundances in their hosts. While some studies did not find a statistically significant difference between these hosts and a general star-forming galaxy population \citep{savaglio09}, others found that LGRB hosts have substantially lower metallicities compared with the field galaxies at the same stellar mass \citep[e.g.,][]{levesque10II,Graham2013}. The observed preference for low metallicities could be a consequence of the occurrence of long GRBs in low mass, actively star-forming galaxies, which dominate the current cosmic SFR (\citealt{mannucci2010,mannucci2011,Campisi2011}; but see \citealt{Graham2013}). Detections of LGRBs in high-metallicity environments suggest that not all LGRB hosts are necessarily characterized by low metallicity and that the metallicity threshold for a GRB production is not as low as predicted by theoretical models \citep{levesque10II,Elliott2013,Kruhler2015,Perley2015b}. Especially at low redshifts ($z < 1.5$) LGRB hosts also differ from other galaxies in properties like star-formation rate (SFR) and stellar mass \citep{Boissier2013,Perley2013,Vergani2014}. To understand the reasons for biased hosts at z $<$ 1.5, more observations of low-z hosts are necessary.

Most of the investigated samples of LGRB hosts are not carefully assembled due to small size or observational limitations. However, \citet{Vergani2014} used a sample which is nearly complete in redshift and unbiased against galaxies in which bright LGRBs exploded \citep{Salvaterra2012, Melandri2012} and showed that $z < 1$ hosts tend to avoid massive galaxies and that this is likely linked to the metallicity of the host. This study showed that, in order to understand the host environment of LGRBs and the reason for the discrepancy between the field galaxies and the LGRB hosts, we should study different aspects of galaxies (e.g, stellar masses, SFRs and metallicities) simultaneously. 

X-Shooter is a second generation instrument mounted at VLT/ESO \citep{Vernet2011}. It has a unique capability to produce intermediate resolution spectra ($R \sim 4000 - 17000$, depending on wavelength and slit width) while simultaneously covering a spectral range from 3000 to 24000 \AA. The instrument enables us to better characterize the LGRB host population and to explore galaxy formation and evolution beyond the parameter space provided by traditional high-$z$ galaxy surveys. 

In this paper we present X-Shooter observations of 6 LGRB host galaxies which lie at $z \sim 1$. The observations were performed during our X-Shooter Guaranteed Time Observation (GTO) program at the VLT/X-Shooter spectrograph\footnote{P.I.'s: H.\ Flores and S.\ Piranomonte from 2009 to 2012}. Some of these hosts have been observed before but lack a good constraint on their metallicity, dust extinction and SFR and some of them have never been observed before. The aim of the survey, and presented study, was to detect emission lines of these LGRB host galaxies in order to measure the mentioned characteristics.

The sample, the observing program and the reduction of the X-Shooter data are presented in Section \ref{reduction}. The measurement procedure of galaxy characteristics is outlined in Section \ref{sec:fluxemission}. A detailed account on results is given in Section \ref{sample} and we discuss them in Section \ref{discuss}. Summary and conclusions are shown in Section \ref{summary}.

\section{Sample, observations and data reduction}
\label{reduction}

\subsection{The sample}
\label{sample1}
In this work we present a study of 6 LGRB host galaxies observed during our GTO program. The hosts of GRB\,000210 \citep{gorosabel03}, GRB\,000418 \citep{gorosabel03b,Hatsukade2011}, GRB\,000911 \citep{price02,masetti05}, GRB\,080413B, GRB\,100316B and GRB\,100724A, were selected because, due to their redshift (in the range $0.8 < z < 1.3$), they were either lacking the measured values of metallicity, dust extinction and SFR or these quantities were very poorly constrained. The spectral coverage and sensitivity of the X-Shooter allows us to get the whole set of lines (\OII, \OIIIb, \Ha, \NII) necessary to perform these measurements. We started our observing program shortly after the end of the X-Shooter commissioning phase, hence three of the observed hosts  (GRB\,000210, GRB\,000418 and GRB\,000911) were selected also because they had previous spectroscopic observations with other telescopes and had one emission line detected (generally \OII) in their spectrum. This would have allowed us to make a comparison of the X-Shooter results with the previous measurements and understand better the instrument performances. For the remaining three galaxies: the host of GRB\,080413B  was observed only in imaging, while the hosts of GRB\,100316B and GRB\,100724A were never observed before. The images we used to build finding charts of these last three hosts (Figure 1), from which spectra we had a lower limit on the redshift, are dominated by afterglow emission. We are aware that this is a very small sample whose implications are discussed in Section \ref{discuss}.

\subsubsection{GRB 000210 host}

GRB\,000210 was discovered by BeppoSAX \citep{piro02}. No optical afterglow was detected down to $R > 23.5$, leading to the classification of this burst as \textit{dark} \citep[e.g.,][]{Jakobsson2004}. The field was observed by the X-ray Chandra telescope 21 hrs after the burst which led to the localization of the X-ray afterglow with an accuracy of 0.6'' radius error circle \citep[][; Figure 1]{piro02}. The $J$-band image of the host shown in Figure \ref{fig:finding} has been obtained with ISAAC at the VLT on Sep 2001 \citep{gorosabel03}.

UV-NIR photometric observations were used to measure the stellar mass of $\approx 10^{9.3}$ \Msun\ \citep{savaglio09,svensson10,Schady2014}.
Including a detection in the sub-mm band, \citet{Michalowski08} found a value of $\sim 10^{10}$ \Msun.

Previous spectral and UV-NIR photometric observations enabled the measurement of the host's SFR $\sim 2$ \Msun\ yr$^{-1}$ \citep{gorosabel03,savaglio09}. A detection of sub-mm emission towards the position of GRB\,000210 was interpreted as originating from the host galaxy and hence suggesting a SFR of several hundred \Msun\ yr$^{-1}$ \citep{berger03,barnard03,Michalowski08}.

\subsubsection{GRB 000418 host}

\begin{figure*}
\centering
{\includegraphics[width=\textwidth]{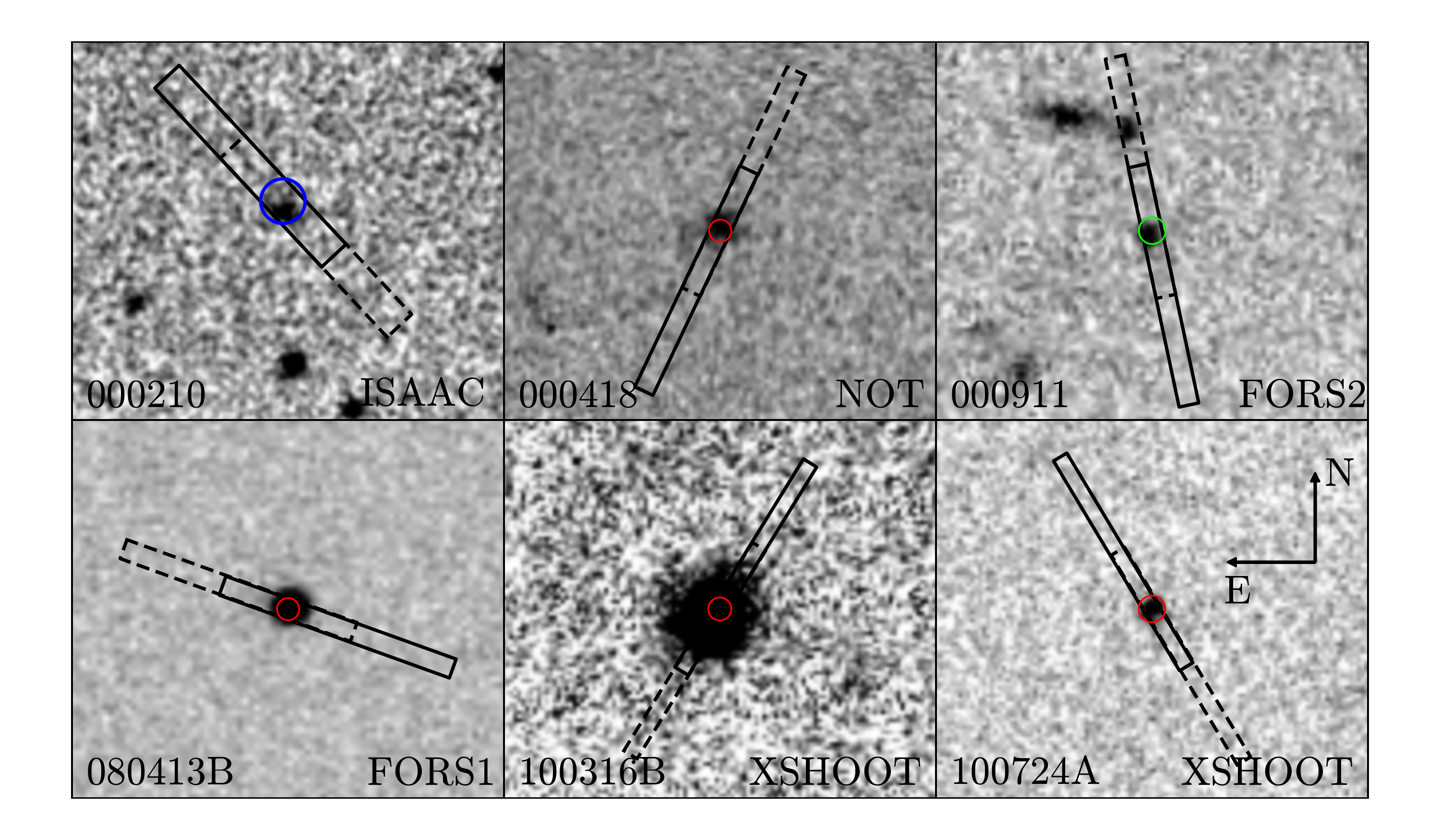}}
\caption{Mosaic of 17$\arcsec\times$17$\arcsec$ regions centered on the positions of the six LGRBs as measured from the afterglow detection. Top three images were obtained several months after the LGRB detection and show the host galaxy. No late-time imaging has been performed for GRBs\,100316B and 100724A and no image of the host of GRB\,080413B is publicly available, therefore the bottom three images show the optical afterglow. Circles indicate the uncertainty of measured X-ray (blue), optical (red) and radio (green) afterglow positions. The placement of the (11$\arcsec$ long) slit as well as the instrument with which the image was obtained are also indicated. All hosts were observed in nodding ABBA sequence with a 5$\arcsec$ offset between A and B slit positions. A and B slit positions are plotted with full and dashed line, respectively. }
\label{fig:finding}
\end{figure*}

The optical afterglow of GRB\,000418 was found to be very red with $R -K \approx 4$ \citep[Vega-calibrated magnitudes;][]{Klose2000}, suggesting a very dusty sight-line environment. A recent modeling of UV-NIR photometric SED, assuming \citet{calzetti00} extinction curve, found an average host extinction of  A$_{V}$= 1.3 mag \citep{Perley2013, Schady2014}. The $R$-band image of the host (Figure \ref{fig:finding}) has been obtained with ALFOSC at the Nordic Optical Telescope approximately one year after the burst \citep{gorosabel03b}.

UV-NIR photometric observations were used to measure the stellar mass of $M_\ast \approx 10^{9}$ \Msun\ \citep{savaglio09,svensson10, Schady2014}. Previous detection of the \OII\ line enabled the measurement of the host's SFR$_{\OII}$ = 55 \Msun yr$^{-1}$. The host was also detected in submillimeter and radio, suggesting higher values of SFR $\sim$ 250-900 \Msun\ yr$^{-1}$ \citep{berger03}.

\begin{table*}
\begin{center}
\small
\begin{tabular}{lcccccccccr}
\hline\hline
LGRB host &  Redshift &  $E_{\rm B -V}$ & Date   & Seeing & Binning & Arm & Exp. Time & Slit width & Resolution \\
            	&                 &               	& Program ID   & &        	& & (s)      	&   ($\arcsec$) 	&  	$\lambda/\Delta\lambda$    	\\
\hline\hline
GRB 000210 &  0.8462 & 0.017 & 20 Nov. 2009   & 1.0" &  1x2  & UVB &1800 x 4  &  1.3   	& 4000     	\\
                	&           	&     	& 084A-0631(A) & &	1x2  & VIS  &1800 x 4  &  1.5   	& 5000     	\\
                  	&             &       	&                        	&  & 1x1   & NIR  & 600 x 12  &  1.5   	& 2600      	\\
                                      	\hline
GRB 000418 & 1.1183 & 0.029 & 16 Apr. 2010   &  0.8" & 1x2  & UVB & 1800 x 2 & 1.0    	& 5400\\
                	&         	&    & 085A-0795(A) &   & 1x2  & VIS  & 1800 x 2 & 0.9    	& 8800\\
                	&         	&    &	                     	&  & 1x1  & NIR & 600 x 6  & 0.9    	& 5800 \\
                                                               	\hline
GRB 000911 & 1.0585 &  0.106 &  21 Nov. 2009   & 1.0"  & 1x2  & UVB & 1800 x 4  &  1.0   	& 5300 \\
                	&         & 	&    	 084A-0631(A) &   & 1x2 & VIS  & 1800 x 4  &  0.9   	& 8500  \\
                  	&          &	&    	                     	&   & 1x1 & NIR &  600 x 12  &  0.9   	& 5700   \\
                        	\hline
GRB 080413B & 1.10124  & 0.032 & 15 Aug. 2010   & 0.6" & 1x2  & UVB & 1800 x 3  &  1.0   	& 4500 \\
                 	&          & 	&      	 085A-0795(B)  & & 1x2  & VIS  & 1800 x 3   &  0.9   	&10000 \\
                 	&         &  	&      	                     	& & 1x1  & NIR &  900 x 3	&  0.9   	&5700 \\
                                                           	\hline
GRB 100316B & 1.181 &   0.124  & 3 Apr. 2011	& 1.3" & 1x2   & UVB & 1200 x 4 & 0.8    	& 6500 \\
                 	&       & 	&        	 087A-0451(A)&  & 1x2   & VIS  & 1200 x 4  & 0.7    	& 11000 \\
                 	&       &	&        	                   	& & 1x1   & NIR  & 400 x 12  & 0.9    	& 5700 \\
                                                            	\hline
GRB 100724A & 1.289 &  0.038  & 3 Apr. 2011	& 0.8" & 1x2  & UVB & 1200 x 4  & 0.8    	& 6500 \\
                 	&       &	&          	 087A-0451(A) & & 1x2  & VIS  & 1200 x 4  & 0.7    	& 11000 \\
                 	&       &	&      	                     	& & 1x1  & NIR &  400 x 12  & 0.9    	& 5700 \\                                                                                                                                                   	 
\hline\hline
\end{tabular}
\caption{
Presentation of the sample and observation log. In columns 1 to 10 we report the GRB name, its host galaxy redshift (measured from emission lines detected in our X-shooter spectra}, see Section \ref{sample}), Galactic color excess, the date of observation together with the observing run, seeing condition, binning, spectrograph arm, total exposure time, slit width and wavelength resolution at the center of each arm, respectively. All observations were performed in nodding mode. We report the average seeing measured by ESO-DIMM \citep{sarazin90} at the time of observations.
\label{tabobs}
\end{center}
\end{table*}

\subsubsection{GRB 000911 host}

Broadband photometric SED modeling of the compact GRB\,000911 host is consistent with a subluminous galaxy  (L $<$ L$_{{\rm \star}}$ ), dominated by an old stellar population, and an average extinction $A_{\rm V} = 0.3-0.5$, assuming a Large Magellanic Cloud \citep{fitzpatrick1986} or starburst galaxy \citep{calzetti00} extinction law \citep{masetti05}. The $I$-band image of the host in Figure \ref{fig:finding} has been obtained with FORS2 at the VLT approximately one year after the burst \citep{masetti05}.

UV-NIR photometric observations were used to measure the stellar mass of  $M_\ast \approx 10^{9.3}$ \Msun\ \citep{savaglio09,svensson10}. Previously detected \OII\, enabled to calculate the SFR$_{\OII} \approx$ 2.2 \Msun\ yr$^{-1}$ \citep{price02b,savaglio09}.

\subsubsection{GRB 080413B host}


To measure the stellar mass of the host of GRB\, 080413B, we use the host observations from GROND \citep{Filgas2011} as well as public \textit{Spitzer}/IRAC data taken as part of the SHOALS survey \citep{Perley2015a,Perley2015b}. IRAC magnitudes were determined based on PBCD data from the archive, circular apertures and tabulated zeropoints. No other galaxies are detected within the immediate vicinity of the host of GRB\, 080413B, and a possible confusion in the IRAC data is no concern in our analysis.

Together, the GROND and IRAC data cover the rest-frame range of 220\,nm to $1.7\,\mu$m and allow us to fit the spectral energy distribution using stellar population synthesis models. Here, we use models from \citet{bruzual2003} and a \citet{Chabrier2003} IMF within Lephare \citep{Arnouts1999, Ilbert2006}. Our input models span a wide range of galaxy parameters as described in detail in \citet{Kruhler2011}. The host's photometric data are well described with a star-forming galaxy template with a $B$-band luminosity of $M_B=-19.4\pm0.3$\,mag and a stellar-mass of $\log\,(M_{\star}/ M_{\odot}) = 9.2_{-0.4}^{+0.2}\,$. Assuming the \citet{Baldry03} IMF in the modeling, values of stellar mass are very similar to \citet{Chabrier2003}, therefore we use this value in further analysis.





\subsection{Observations}

We obtained X-Shooter spectra of the 6 LGRB hosts in the period between November 2009 and April 2011. Slit positions with respect to the positions of the LGRB afterglows are shown in Figure \ref{fig:finding}. All spectra were taken under dark sky, but different seeing conditions. In each case the individual exposures were obtained by nodding along the slit with an offset of 5" between exposures in a standard ABBA sequence. Only two exposures (AB) were acquired in the case of GRB\,000418 and only three (ABB) in the case of GRB\,080413B due to clouds during both observations. Observation logs are given in Table \ref{tabobs}.

 \subsection{Data reduction}

We processed the spectra using version 2.0.0 of the X-Shooter data reduction pipeline \citep{Goldoni06, Modigliani2010}, combined with the reduction technique developed for NODDING mode observations. The pipeline transforms the detected counts in physical quantities while propagating the errors and bad pixels consistently. First, the raw frames are subtracted from each other and cosmic rays are detected and removed using the method developed by \citet{vanDokkum01}. The frames were divided by master flat-fields produced during day time with halogen lamps. Orders were extracted and rectified in wavelength space using the wavelength-spatial solution previously obtained from calibration frames. The resulting rectified frames were then shifted by the offset (5$\arcsec$), added and merged, thus the final 2D spectrum was obtained. The merging in the overlapping regions was weighted by the errors. The 1D spectrum together with the corresponding error file and bad pixel map, extracted at the source position from the 2D spectrum, are the final products of the pipeline. More than one source is found in the slit in the case of GRB\,000911, therefore the reduction for this host was done separately for each nodding step using a STARE mode reduction. Individual extracted 1D products were averaged to obtain the final 1D spectrum.

The flux calibration of the spectra is particularly challenging for an instrument such as X-Shooter, which combines together three different detectors in UV, optical and NIR (e.g., see  \citealt{Kruhler2015} and \citealt{Japelj2015}). First, we used observations of a spectrophotometric standard star, taken  on the same night as science spectra, to flux calibrate the spectra. The reduced standard star spectrum was divided by the tabulated spectrum to produce the response function: the response was interpolated in the atmospheric absorption regions in VIS and NIR spectra. The response function was then applied to the spectrum of the host. Flux values were subsequently cross-checked and corrected using the corresponding magnitudes of the galaxies. This was possible for GRB\,000210, 000418 and 000911 hosts according to the photometric studies published in the literature\footnote{Published measurements were mainly taken from the public database GHostS at www.grbhosts.org.}. Slit-aperture flux loss corrections are therefore included in the procedure.

For the cases where the photometry is not available or the continuum is not detected, we checked the validity of flux calibration with telluric stars. Spectra of telluric stars were taken right after host observations and with the same instrumental setup (binning, slit width). Telluric stars were observed in the nodding mode and the reduction procedure was the same as for the host spectrum. The same correction factors, obtained by comparing flux calibrated spectra of telluric stars to their magnitudes, were then applied to the spectra of the hosts. The method was tested in cases where magnitudes are available for both telluric stars and hosts: the uncertainty of this calibration (which does not include slit loss correction) is of the order of 20$\%$. This method was introduced and validated by \citet{Pita2014} who find similar conclusions. Due to the large uncertainty we do not include slit-aperture flux loss corrections. The method was used to cross-calibrate the full spectra of GRB\,080413B, 100316B and 100724A hosts.

\begin{table*}
\begin{center}
\begin{tabular}{lcccccc}
\hline\hline
Line & GRB\,000210 & GRB\,000418 &GRB\,000911 & GRB\,080413B & GRB\,100316B $^{(b)}$ & GRB\,100724A \\
\hline\hline

\OII$\lambda3726^{(a)}$ & $7.4\pm0.4$ & $3.6 \pm 0.6$ &$0.5\pm0.1$ & $0.8\pm0.1$ & $5.6\pm0.4$ & $ 1.0 \pm 0.1$ \\
\OII$\lambda3729$ & & $5.1 \pm 0.6$ &$0.9 \pm 0.1$ & $1.0 \pm 0.1$ & & $1.9 \pm 0.1$ \\
\NeIII$\lambda3869$ & $0.3\pm0.2^{(c)}$ & $<1.1$ & $< 0.4$ & $0.2 \pm 0.1$ & $< 0.7$ & $< 0.7$ \\
H$\gamma$ & $0.8\pm0.3^{(c)}$ & $0.8 \pm 0.3^{(c)}$ &$< 0.4$ & $0.2 \pm 0.1^{(g)}$ & $0.6\pm0.4^{(h)}$ & $< 1.5$ \\
H$\beta$ & $2.2\pm0.3$ &$<7.2$ & $<1.1$ &$<2.2$ & $<1.3$ & $<1.3$ \\
\OIIIb$\lambda4959$ &$1.2 \pm 0.3$ &$5.4 \pm 1.8$ &$<3.4$ &$^{(e)}$ & $2.6\pm0.8$ & $^{(e)}$ \\
\OIIIb$\lambda5007$ &$4.7 \pm 0.3$ &$11.4 \pm 1.8$ &$<3.4$ &$2.2 \pm 0.3$ & $^{(i)}$ & $2.8 \pm 0.5$ \\
H$\alpha$ & $ 6.1 ^{(d)}$ & $^{(f)}$ & $^{(f)}$ &$1.4^{(d)}$ & $^{(f)}$ & $3.1 \pm 0.4$\\
\NII$\lambda6584$ &$^{(e)}$ & $^{(f)}$ & $^{(f)}$ &$^{(f)}$ & $^{(f)}$ & $^{(e)}$ \\

\hline\hline
\end{tabular}
\caption{Emission-line fluxes (in units of $10^{-17}$\,erg s$^{-1}$ cm$^{-2}$) are corrected for Galactic extinction and stellar Balmer absorption  (see Section \ref{sec:fluxemission} for details). Correction due to dust extinction in the host galaxies has not been applied. We report 3-sigma upper limits when lines are not detected. \newline
(a) If the \OII\ doublet is clearly resolved, we report the flux for each line separately;
(b) Fluxes for the 'host' component (see Section \ref{XS100316B} for details);
(c) Uncertain flux, but line is clearly detected in 2D;
(d) See section 4 for details on the flux measurements;
(e) Undetected line falling on a sky line;
(f) Line falling in a telluric region;
(g) There is a clear detection in 2D at the corresponding wavelength but the shape is different from the other lines;
(h) Line not detected in 2D.
(i) Line partially detected but strongly affected by a sky line.
}

\label{tabfluxes}
\end{center}
\end{table*}

\section{Measurements}
\label{sec:fluxemission}

The emission line fluxes were measured using the code developed by us and cross-checked with the \textit{splot} task in IRAF.\footnote{IRAF is software distributed by \textit{National Optical Astronomy Observatories}.} In cases where a line is not detected we report a 3$\sigma$ upper limit. All fluxes are corrected for Galactic extinction in the direction of the host assuming the extinction curve of \citet{Cardelli1989} and reddening values given in Table \ref{tabobs} \citep{Schlafly2011}. Throughout the paper, we report line fluxes in units of $10^{-17}$ erg \, cm$^{-2}$ s$^{-1}$.

Galaxies hosting young or intermediate age stellar populations feature strong Balmer absorption lines in their spectra. Thus the Balmer emission lines are superimposed on the host stellar absorption lines (the absorption is typically negligible for metal lines). Due to a faint spectral continuum we cannot detect the wings of absorption lines and therefore cannot correct for stellar absorption directly. \citet{Zahid2011} measured the average equivalent width ($EW$) of H$\beta$ absorption line as a function of stellar mass and spectral resolution for a population of blue, low-mass galaxies, which is similar to the LGRB hosts of our sample. Given the X-Shooter resolution (Table \ref{tabobs}) the results of \citet{Zahid2011} imply a typical rest-frame equivalent width of stellar H$\beta$ absorption line of $1 \mathrm{\AA}$. We adopted this value to estimate the correction for the stellar absorption. The correction was non-negligible only for the H$\beta$ line of GRB\,100316B, where we estimate it to be F$_{\rm corr} = F_{\rm cont}(1 + z)1\mathrm{\AA} \sim 0.2\times10^{-17}$ erg cm$^{-2}$ s$^{-1}$.

Line fluxes, corrected for Galactic extinction and stellar absorption are reported in Table\,\ref{tabfluxes}.

Dust extinction in the hosts is measured by comparing Balmer line ratios to theoretical expectations, calculated for nebular temperature of $10^{4}$ K \citep[$F_{\mathrm{H}\alpha}/F_{\mathrm{H}\beta} = 2.87$, $F_{\mathrm{H}\gamma}/F_{\mathrm{H}\beta} = 0.47$;][]{osterbrock06}. 
In the cases where Balmer lines are not detected, we searched through the literature for any estimate of dust extinction either from emission lines in the host-integrated signal or line-of-sight (via SED analysis of the afterglow). The details of host extinction correction for each host in our sample are given in Section \ref{sample} and the final adopted values are reported in Table \ref{tabmetalsfr}. Dust-corrected emission-line fluxes are used in the analysis.

\subsection{Star-formation rates}
\label{sfr}

Optical nebular lines effectively originate from the re-emission of stellar luminosity in galaxies at wavelengths shortward of the Lyman limit, therefore directly probe the young, massive stellar populations \citep{Kennicutt98}. We estimated the SFR by using emission luminosities $L$ from dust-extinction and stellar-absorption corrected H$\alpha$, H$\beta$ and \OII\ line fluxes and the following prescriptions:

\begin{equation}
\label{eq:SFRHa}
SFR_{\rm H\alpha} = 4.39 \times 10^{-42} \frac{L_{\rm H\alpha}}{\mathrm{erg} ~\mathrm{s}^{-1}}\Msun\ yr^{-1},
\end{equation}
\begin{equation}
\label{eq:SFRHb}
SFR_{\rm H\beta} = 12.6 \times 10^{-42} \frac{L_{\rm H\beta}}{\rm erg ~s^{-1}}\Msun\ yr^{-1},
\end{equation}
\begin{equation}
\label{eq:SFROII}
SFR_{\OII} = 5.54 \times 10^{-42} \frac{L_{\OII}}{\rm erg ~s^{-1}}\Msun\ yr^{-1},
\end{equation}

\noindent
where the relation for H$\alpha$ line is the one reported by \citet{Kennicutt98}, but divided by a factor of 1.8 to go from Salpeter IMF \citep{Salpeter1955} to a more realistic IMF \citep{Baldry03}. SFR for H$\beta$ is obtained by using the Balmer decrement at typical nebular temperatures of $10^{4}$ K (H$\alpha$/H$\beta$ = 2.87). The \OII\, SFR relation is the one empirically derived specifically for LGRB hosts by \citet{savaglio09}. The advantage of using the \OII\ as a tracer of SFR is that this line is usually stronger than H$\beta$ and easy to detect. However, it is more affected by dust extinction. Moreover, [OII] depends on the metallicity and ionization level. Nevertheless, \citet{savaglio09} showed that it can still be quite reliable and thus a valuable tool in case the two hydrogen lines are contaminated or undetected.

We measure SFRs for all six hosts in the sample. SFR values are reported in Table \ref{tabmetalsfr}.

\begin{table*}
\renewcommand{\arraystretch}{1.3}
\begin{center}
\scriptsize
\begin{tabular*}{\textwidth}{l @{\extracolsep{\fill}} ccccccccccr}
\hline
& & \multicolumn{5}{c}{$12+\log (\mathrm{O/H})^{(a)}$} & \multicolumn{3}{c}{SFR (\Msun yr$^{-1}$)} &  \\
\cmidrule(lr{.75em}){3-7}
\cmidrule(lr{.75em}){8-10}
Host &A$_{V}$ & M91 & KK04 & P05 & \NeIII/\OII & M08 & H$\alpha$ & H$\beta$ & $\OII$ & $n_{\rm e}$ (cm$^{-3}$) &  log (M$_{*}$/M$_{\odot}$) \\
\hline
000210 & 0 & $8.1 \pm 0.1$ & $8.2 \pm 0.1$ & $7.9 \pm 0.2$ & $8.7 \pm 0.2$ & $8.6 \pm 0.1$ & $\sim$ 0.9 & 1.0$\pm$0.1 & 1.4$\pm$0.1 & & 9.31$_{-0.08}^{+0.08}$\\
& & $8.6 \pm 0.1$ & $8.7 \pm 0.1$ & $8.2 \pm 0.1$ & & & & & \\
000418 & 1.3 & $\sim 8.4$ & $\sim 8.4$ & & $> 8.3$ & $> 8.3$ & & $< 25.4$ & 19.0$\pm$1.9 & $< 100$ & 9.26$_{-0.14}^{+0.14}$\\
000911 & 0.4 & & & & $> 7.9$ & 8.1$\pm$0.6 & & $< 1.3$ & 0.8$\pm$0.2 & $< 10$ & 9.32$_{-0.26}^{+0.26}$\\
080413B$^{(b)}$ & 0 & $\sim 8.4$ & $\sim 8.4$ & $\sim 8.4$  & $8.4 \pm 0.2$ & $8.4 \pm 0.2$ &$0.4 \pm 0.1$ & $<1.9$ & 0.7$\pm$0.1 & $< 30$ & 9.2$_{-0.4}^{+0.2}$\\
100316B & $0$ & $\sim 8.4$ & $\sim 8.4$ & & $>8.3$ &  $>7.9$ & & $<1.3$ & 2.5$\pm$0.2& &\\
100724A & $0$ & $8.0 \pm 0.1$ & $8.2 \pm 0.1$ & $7.9 \pm 0.2$ & $>8.0$ & & $1.4 \pm 0.2$ & $< 1.6$ & $1.6\pm 0.1$ & &\\
& & $8.6 \pm 0.1$ & $8.8 \pm 0.1$ & $8.3 \pm 0.1$ & & & & & & &\\
\hline
\end{tabular*}
\caption{Host parameters measured from detected lines (Table \ref{tabfluxes}): host extinction, metallicites from different calibrations, SFR and electron density. Measurement details for each specific galaxy are given in Section \ref{sample}. In the last column we add values of hosts' stellar masses, when  available.\newline
(a) M91, KK04 and P05 metallicities are based on the degenerate $R_{23}$ calibration, thus both solutions are reported. Metallicity errors do not include calibration uncertainties (see text); (b) the theoretical Balmer decrement $F_{\mathrm{H}\beta}$ = $F_{\mathrm{H}\alpha}$/2.87 is assumed; 
}
\label{tabmetalsfr}
\end{center}
\end{table*}

\subsection{Metallicities}
\label{metal}

Emission lines in HII regions have been used for a long time to derive metallicities in the ionized interstellar medium of galaxies. A number of metallicity calibrations have been developed in the past three decades (see \citealt{Kewley08} for a review). The calibrations were derived either with the help of theoretical models or from empirical relations or from a combination of the two. In all cases the metallicity can be estimated by using relations that are a function of different combinations of emission-line flux ratios. Unfortunately, each method is affected by systematic errors, which may also be function of redshift (ionization is perhaps the evolving physical parameter). When comparing the metallicity of different objects it is therefore important to measure the metallicities with the same calibration method.


According to the line set available, we can use two calibrations which are based on theoretical photoionization models (M91 - \citealt{McGaugh1991}; KK04 - \citealt{Kobulnicky2004}) and on two empirical calibrations (P05 - \citealt{Pilyugin2005}; \NeIII/\OII\ - \citealt{maiolino08}). M91, KK04 and P05 calibrations rely on the so-called $R_{23}$ parameter, defined as:

\begin{equation}
\label{eq:R23}
R_{23}=\frac{\OII\lambda3727 + \OIIIb\lambda4959 + \OIIIb\lambda5007}{H\beta},
\end{equation}

where the $\OII\lambda3727$ is the total flux of the doublet. All these lines are usually strong and easily detectable. The calibrations could have problems if the host extinction is not properly taken into account. The $R_{23}$ has one major drawback: it gives two solutions for the metallicity, the upper and the lower branch solution. The most probable metallicity can be chosen by cross correlating with additional information, for instance if the \NII/\OII\ or \NII/H$\alpha$ ratio - both strong functions of metallicity \citep{Kewley02,Kewley08} - is measured. In our sample, these two ratios are not available. In the KK04 calibration, an additional parameter considered in the $R_{23}$ relation is the ionization defined by the line ratio $O_{32} = \mathrm{\OIIIb}\lambda 4959,$ $5007/\mathrm{\OII\lambda3727 }$. In the KK04 calibration, the ionization parameter in the lower- and upper-branch solution is obtained iteratively. The O32 ratio itself can be used to choose the branch of $R_{23}$ \citep{nagao06,maiolino08}. The P05 calibration includes the excitation parameter $P = (\mathrm{\OIIIb}\lambda 4959,$ $5007/\mathrm{H}\beta)/R_{23}$. Another useful calibration uses the line ratio \NeIII$\lambda 3869$/\OII$\lambda3727$ \citep{maiolino08}. This ratio has the advantage that the lines are very close, and the dust correction has a minimal effect, therefore it is a valuable method when extinction is unknown. Moreover, in contrast to the other three methods used in this work, its solution is unique (but it is sensitive to the ionization parameter). Uncertainties for M91, KK04, P05 and \NeIII/\OII\ calibrations are estimated to be 0.15, 0.15, 0.1 and 0.15 dex, respectively. All measured metallicities are summarized in Table \ref{tabmetalsfr}.

We also measure metallicities by simultaneously considering all flux ratios between relevant emission lines and fitting them with two free parameters: metallicity and host extinction \citep[M08;][]{maiolino08}. In the case of a good dataset this method can surpass problems faced by individual line ratios (e.g., double valued solution of R$_{23}$ calibrator or poorly known extinction). Most of the relevant lines have been measured for the GRB\, 000210 and 080413B hosts, for which we measure a negligible host extinction and metallicities of 12+log(O/H) of $8.6 \pm 0.1$ and $8.4 \pm 0.2$, respectively. All the other hosts lack detection of the H$\beta$ line, and in the two cases where the H$\alpha$ is detected, \NII$\lambda6584$ is not. This makes the measurement of metallicity problematic. Limits or rough estimates can be obtained from O32 for GRB000418 and GRB100724A and from upper and lower limits to flux ratios for the others.

\subsection{Electron densities}
\label{dense}

The doublet \OII$\lambda3726,3729$ line ratio is sensitive to the electron density in the gas $n_e$ \citep{osterbrock06}. For those hosts with resolved \OII\ lines, the electron density is estimated by assuming a typical temperature in the star-forming region $T \sim 10^{4}$ K (Figure 5.8 in \citealt{osterbrock06}). We were only able to measure upper limits on $n_e$: the results are shown in Table\,\ref{tabmetalsfr}.

\section{Results}
\label{sample}

In the following we present detailed results for each individual host discussing both the characteristics already published and the new X-Shooter results. As an example of X-Shooter spectra, we show the figures of 1D and 2D spectra in the proximity of detected lines for GRB\,000210 host. SFRs and stellar masses are reported assuming a \citet{Baldry03} IMF. 

\subsection{GRB\,000210 host}
\label{000210}
\begin{figure}
\centering
{\includegraphics[width=\linewidth]{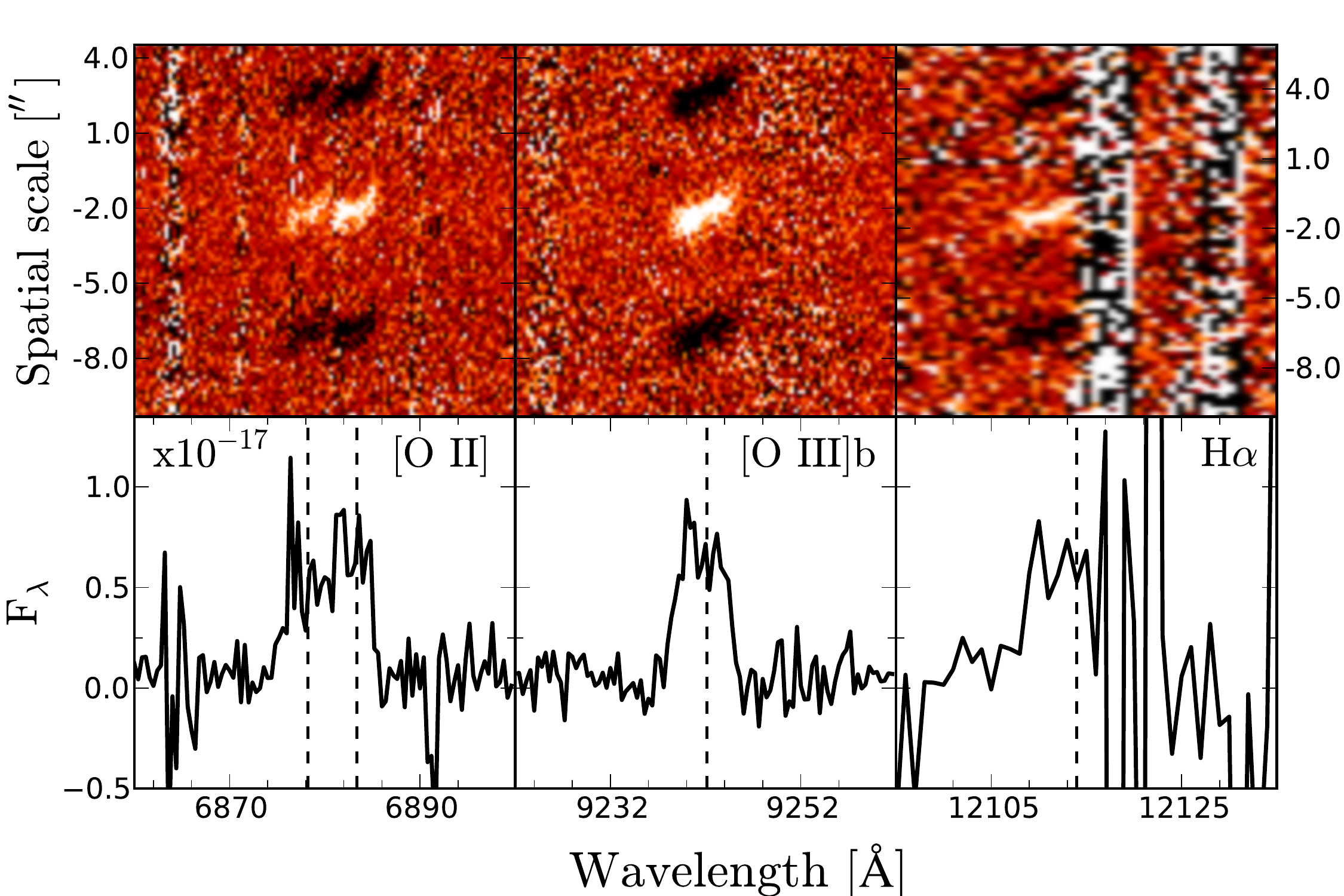}}
\caption{ \OII, \OIIIb$\lambda 5007$ and H$\alpha$ lines as observed by the X-Shooter for GRB\,000210 host. Both 2D (up) as well as extracted 1D (bottom) spectra are plotted for each line. In general, spatial and wavelength scale of VIS and NIR differ. Vertical dashed lines show redshifted air wavelengths of corresponding lines. Flux calibration of 1D spectra is not scaled with respect to photometric measurements or telluric stars (see Section \ref{reduction}).}
\label{fig:OIII_000210}
\end{figure}

\begin{figure}
\centering
{\includegraphics[width=\linewidth]{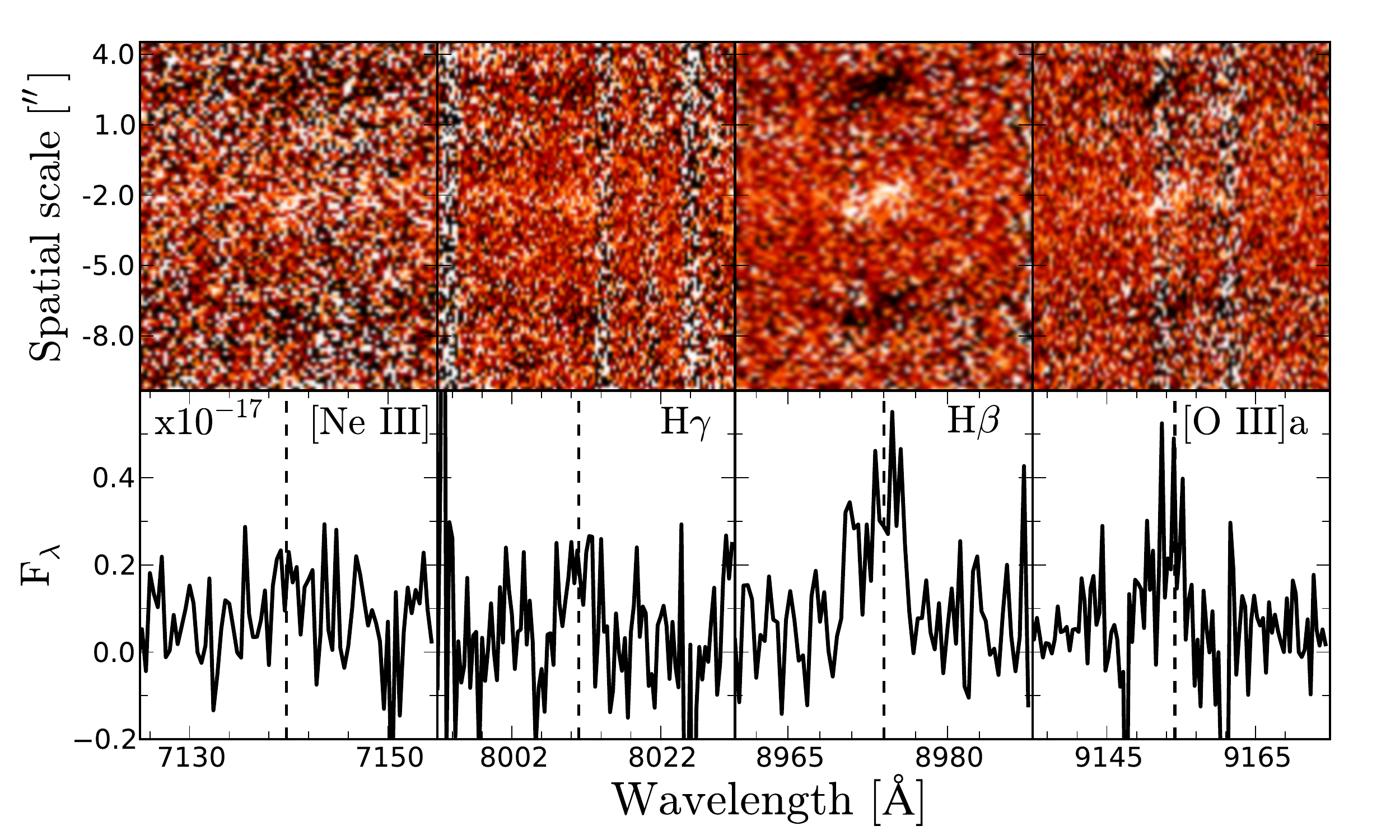}}
\caption{\NeIII, H$\gamma$, H$\beta$, and \OIIIb$\lambda 4959$ lines as observed by the X-Shooter for GRB\,000210 host.}
\label{fig:Hb_000210}
\end{figure}

The X-Shooter spectrum of the GRB\,000210 host galaxy reveals several emission lines: \OII, \NeIII, H$\gamma$, H$\beta$, \OIIIb\ and H$\alpha$ ( Figures \ref{fig:OIII_000210} and \ref{fig:Hb_000210}). The H$\alpha$ line is marginally contaminated by a sky line. Considering the shape and the velocity spread (about 250\,km\,s$^{-1}$) of the lines, we could estimate the amount of missing flux and apply the correction. The derived values are consistent with theoretical H$\alpha$/H$\beta$ and H$\alpha$/H$\gamma$ ratios, expected for a negligible (within errors) dust extinction. Negligible dust extinction is also derived by the modeling of the host SED with synthetic galaxy spectra \citep{gorosabel03}. GRB\,000210 was classified as dark and even though a typical dark GRB host is significantly dust obscured, there are known cases of dark GRB hosts with very low or even negligible host-averaged dust extinction \citep{Perley2013, Kruhler2011}.

H$\alpha$, H$\beta$ and \OII\ line fluxes give SFR $=\sim$0.9, $1.0\pm0.1$ and $1.4\pm0.1$ \Msun\ yr$^{-1}$, respectively. These values are a factor $\sim 2$ lower than the ones reported by \citet{savaglio09}. The discrepancy is due to the average dust extinction correction (with $A_{\rm V} = 0.53$) applied by \citet{savaglio09} for all hosts in their sample which had no dust extinction estimate. On the other hand the UV continuum-based SFR obtained by \citet{savaglio09} SFR$_{2800} \approx 1.34$ \Msun\ yr$^{-1}$, is similar to our results.

R$_{23}$-based metallicities are measured to be $8.1\pm0.1$, $8.2\pm0.1$, $7.9\pm0.2$ for lower-branch, and $8.6\pm0.1$, $8.7\pm0.1$, $8.2\pm0.1$ for upper-branch solution by using the M91, KK04, P05 calibrators, respectively. Since \NII\ is not detected, we cannot break the degeneracy. \citet{nagao06} and \citet{maiolino08} showed that, in the absence of \NII, the \OIIIb$\lambda5007$/\OII$\lambda3727$ ratio could be used instead. We measure \OIIIb$\lambda5007$/\OII$\lambda3727 \sim 0.7$, which together with the \NeIII$\lambda 3869$/\OII$\lambda3727$ ratio are only consistent with the upper-branch solution. 
Additionally, the rich line dataset  allows us to use the M08 method to measure metallicity and extinction of the host: the best-fit parameters of the multiple line-ratios fit are Z$_{\rm M08} = 8.6 \pm 0.1$ and A$_{\rm V} \sim 0$.

\OII\ doublet is barely resolved (Fig. \ref{fig:OIII_000210}), therefore we do not attempt to measure the average electron density in H II regions.

Using the SFR from H$\alpha$ and a stellar mass of $M_\ast \sim 10^{9.31}$ \Msun\ from \citet{savaglio09}, we calculate a specific star-formation rate SFR/$M_\ast =$sSFR $\approx$ 0.4 Gyr$^{-1}$, which is consistent with the one derived for other LGRB hosts at $z \sim 1$ \citep{Hunt2014}.


\subsection{GRB\,000418 host}

We detect four emission lines with X-Shooter: \OII, H$\gamma$ and the \OIIIb\ doublet spanning a velocity range of about 190\,km\,s$^{-1}$. We do not detect H$\alpha$ and only poorly constrain the upper limit for H$\beta$ (which falls on the edge of the NIR arm), therefore we cannot use Balmer line ratios to determine host extinction. To correct our flux values for host extinction, we adopt the value of $A_{\rm V} = 1.3$, which is the average host optical extinction found by \citet{Perley2013} and \citet{Schady2014} (similar extinction of $0.96 \pm 0.20$ has been measured in the GRB line-of-sight by \citealt{Klose2000}).

\OII\ gives SFR $= 19.0\pm1.9$ \Msun yr$^{-1}$ (consistent with the H$\beta$ upper limit). The lower limit to \OII/H$\beta$ and upper limit to \NeIII/\OII\ together with a measure of \OIIIb/\OII, gives a metallicity $12+\log (\mathrm{O/H}) > 8.2$ with the M08 method. For $R_{23}$ metallicities, we use H$\gamma$ to estimate $F_{\mathrm{H}\beta}$ = $F_{\mathrm{H}\gamma}$/0.47 $= (2.7 \pm 1.0)\times 10^{-17}$ erg s$^{-1}$ cm$^{-2}$. In the case of M91 and KK04 calibrators, the $R_{23}$ value is at the turnover point ($\log R_{23} = 1$; \cite{Kobulnicky2004}) meaning that the metallicity is $12+\log(O/H) \sim 8.4$, equivalently for M91 and KK04 calibrators. In the case of P05, the solution is unphysical ($\log R_{23} > 1$). We note that H$\gamma$ is only tentatively detected (2.7$\sigma$ significant) and the measured flux highly uncertain.

Taking errors into account, we conservatively measure the electron density for this host as $n_{e} < 100$ cm$^{-3}$.
Using our measured value of SFR$_{\rm \OII}$ and a stellar mass $M_\ast \sim 10^{9.26}$ \Msun\ \citep{savaglio09}, we calculate sSFR $\approx$ 3.1 Gyr$^{-1}$.

\subsection{GRB\,000911 host}

We could identify in the spectra three emission line sets: one of the host galaxy at $z=1.058$, one at $z=0.646$ at a position about 4.5\arcsec\, NE of the host (\OII, H$\beta$, \OIIIb, H$\alpha$) and a third one at $z=0.823$, about 1.8\arcsec\, SW from the host, for which only the \OIIIb\, doublet is detected. The galaxy at $z=0.646$ would contaminate the host spectrum in a NODDING mode reduction, therefore we reduced the host spectrum in a STARE mode.

Only the \OII\ doublet is detected for the LGRB host, spanning about 130\,km\,s$^{-1}$. H$\alpha$ falls in a telluric absorption band, and H$\beta$ and \OIIIb\, are in a very noisy part at the beginning of the NIR spectrum, therefore we can determine only flux limits. To correct the \OII\ fluxes for host extinction, we use $A_{V} = 0.4$ and a Large Magellanic Cloud extinction curve \citep{masetti05} - the value is consistent within errors with the GRB line-of-sight host extinction measured to be $A_{\rm V} = 0.27 \pm 0.32$ \citep{Kann2006}.

We measured SFR$_{\OII} = 0.8 \pm 0.2$ \Msun yr$^{-1}$, which is lower than previous measurements based on the \OII\ line, but in agreement with the UV continuum-based SFR$_{2800}$ found by \citet{savaglio09}. Due to the sparsity of detected lines and the limits found on \OIIIb/\OII, \OII/H$\beta$ and \NeIII/\OII\ we were only able to determine a lower limit to the metallicity $12+\log (\mathrm{O/H}) > 7.9$ based on the \NeIII/\OII\ calibration method, while the M08 method resulted in an unconstrained valued.

The electron density is measured to be $n_e < 10$ cm$^{-3}$. Using our measured SFR and a stellar mass $M_\ast \sim 10^{9.32}$ \Msun\ \citep{savaglio09}, we calculate sSFR$_{\OII}$ $\approx$ 0.4 Gyr$^{-1}$.

\subsection{GRB\,080413B host}

Five emission lines are detected in the X-Shooter spectrum of the GRB\,080413B host: \OII\ doublet, \NeIII, H$\gamma$, \OIIIb$\lambda 5007$ and H$\alpha$ spanning about 130\,km\,s$^{-1}$. Lines are found to originate at a common redshift $z = 1.10124$, consistent with the redshift $z = 1.1014$ found from the afterglow spectrum \citep{fynbo09}, therefore confirming association of detected emission lines with the GRB\,080413B host galaxy.

The H$\alpha$ line is detected, but strongly affected by the telluric absorption. We used the spectrum of the telluric star observed right after the host to extract the telluric spectrum and correct the region around H$\alpha$ (the corrected flux is reported in Table 2). We caution that the spectrum of the telluric star was taken at different airmass than the spectrum of the host, adding some extra uncertainty to the reported flux. We note that the measured value $F_{\mathrm{H}\alpha} \approx 1.4\times 10^{-17} $ erg s$^{-1}$ cm$^{-2}$ is very close to the one expected from the detected H$\gamma$ (i.e., $F_{\mathrm{H}\alpha} \approx 6\times F_{\mathrm{H}\gamma} \approx 1.2 \times 10^{-17} $ erg s$^{-1}$ cm$^{-2}$).

The dust extinction could not be determined from the Balmer decrement. \citet{Filgas2011} studied the afterglow of GRB 080413B and found that the sight-line host extinction is negligible. Thus, we do not apply any dust extinction correction.

\OII\ and H$\alpha$ lines are used to measure SFR $= 0.7\pm0.1$ \Msun yr$^{-1}$ and SFR $= 0.4\pm0.1$ \Msun yr$^{-1}$, respectively. The detection of \NeIII, \OII\ and \OIIIb\ enables us to measure the M08 metallicity $12+\log (\mathrm{O/H}) = 8.4 \pm 0.2$. Since the weakest line of the [OIII] is not detected, we assume the theoretical line ratio and estimate $F_{\mathrm{\OIIIb}\lambda 4959} = F_{\mathrm{\OIIIb}\lambda 5007}/3 = 0.7 \pm 0.3$ \citep{storey00}. We infer H$\beta$ flux from H$\alpha$: $F_{\mathrm{H}\beta}$ = $F_{\mathrm{H}\alpha}/2.87 \approx 0.5\times 10^{-17} $ erg s$^{-1}$ cm$^{-2}$. According to all three $R_{23}$ calibrators the solution is consistent, within errors, with the turnover point at $\log R_{23} = 1$ \citep{Kobulnicky2004}, that is, $12+\log (\mathrm{O/H}) \sim 8.4$ which is very close to the measured M08 metallicity.

The electron density is estimated to be $n_e < 30$ cm$^{-3}$. Using our measured SFR and stellar mass we calculate sSFR$_{\OII}$ $\approx$ 0.4 Gyr$^{-1}$.

\subsection{GRB 100316B host}
\label{XS100316B}

In the X-Shooter spectra of the host, a strong \OII\ doublet spreading more than 450\,km\,s$^{-1}$ is clearly detected at the afterglow coordinates. Its profile is asymmetric. There is also a continuum trace clearly detected all along the spectra, centred about 1\arcsec\ below the afterglow coordinates. Due to the lack of deep images of the field and poor seeing conditions during observations, it is not possible to determine if the line emission and continuum belong to the same system. Nonetheless, the spatial profile of the \OII\ doublet in the 2D spectrum seems to indicate a second smaller peak at the continuum centre, suggesting the presence of one single galaxy with a particularly active region, or two close galaxies at the same redshift one of which faint and star forming (that at the afterglow position) and the other brighter and less active. Deep observations of the host and ideally high spatially resolved IFU spectroscopy could give information on this interesting system.

For our analysis we extracted two sets of 1D spectra: one centered on the \OII\ doublet detected at the afterglow position (called here HOST and whose fluxes are reported in Table 2) and with a spatial coverage as large as the seeing (1.3\arcsec), and the other including also the continuum region (hereafter 'TOT'). For the HOST system we detect also \OIIIb. The noise in the TOT spectra is much higher and we can report only limits for this line. No other lines are detected. The \OII\ doublet flux in the TOT extraction is $8.1\pm0.5\time10^{-17}$\,erg s$^{-1}$ cm$^{-2}$.

\OII\ doublet, H$\gamma$ and \OIIIb\ lie at a common redshift of $z = 1.181$, consistent with the redshift $z = 1.180$ found for the afterglow spectroscopy \citep{vergani10}.

\citet{Greiner2011} studied the afterglow of GRB 100316B and found that the sight-line host extinction is negligible. A negligible extinction is also inferred from the detected H$\gamma$ and H$\beta$ upper limit. We thus assume $A_{\rm V} \sim 0$ for the host extinction.

From the \OII\ emission line we measure a HOST SFR$_{\OII} = 2.5\pm0.2$ \Msun yr$^{-1}$. We were able to determine a lower limit to the metallicity $12+\log (\mathrm{O/H}) > 8.3, 7.9$ based on the \NeIII/\OII, M08 calibration, respectively. Even though H$\gamma$ is only tentatively detected, we can use it to estimate the metallicity from $R_{23}$ by assuming $F_{\mathrm{H}\beta} = F_{\mathrm{H}\gamma}/0.47 \sim 1.3$. We also assume F$_{\mathrm{\OIIIb}\lambda 5007}= 3\times F_{\mathrm{\OIIIb}\lambda 4959} = 7.8\pm2.3$. In the case of M91 and KK04 calibrators, the solution is consistent, within errors, with the turnover point at $\log R_{23} = 1$ \citep{Kobulnicky2004}, that is, $12+\log (\mathrm{O/H}) \sim 8.4$. For the P05 calibration, the solution is unphysical ($\log R_{23} > 1$).

The \OII\ doublet is not resolved, therefore we cannot measure the average electron density in H II regions.

\subsection{GRB 100724A host}

Three emission lines are detected in the X-Shooter spectrum: the \OII\ doublet, \OIIIb$\lambda 5007$ and H$\alpha$, spanning about 200\,km\,s$^{-1}$. Lines are found to originate at a common redshift $z = 1.289$, consistent with the redshift $z = 1.288$ found from the afterglow spectroscopy \citep{thoene10}. Even though the continuum is not detected in the NIR arm, upper limits on the continuum are so low that the correction to the measured flux values is completely within the estimated errors. H$\beta$ and H$\gamma$ are too loosely constrained to provide any constraint on the host extinction. Since there are no other observations and data on the host dust extinction in the literature, we neglect the dust extincion.

H$\alpha$ and \OII\ lines are used to estimate SFR $=1.4\pm0.2$ \Msun yr$^{-1}$ and SFR $=1.6\pm0.1$ \Msun yr$^{-1}$, respectively. Because we do not know the extinction in the host, these values are formally  lower limits. We were able to determine a lower limit to the metallicity $12+\log (\mathrm{O/H}) > 8.0$ based on the $\NeIII/\OII$ calibration. For $R_{23}$, we assume $F_{\mathrm{\OIIIb}\lambda 4959} = F_{\mathrm{\OIIIb}\lambda 5007}/3 = 0.9\pm0.2$ and $F_{\mathrm{H}\beta}$ = $F_{\mathrm{H}\alpha}$/2.87 = $1.1\pm0.1$. Thus the R$_{23}$-based metallicities are $8.0\pm0.1$, $8.2\pm0.1$ and $7.9\pm0.2$ for the lower-branch solution, and $8.6\pm0.1$, $8.8\pm0.1$ and $8.3\pm0.1$ for the upper-branch solution, for M91, KK04 and P05, respectively. We note that the value of H$\beta$ obtained in this case is much more reliable than the one inferred from H$\gamma$ for other hosts, because H$\alpha$ is detected with high significance.

Even though the \OII\ doublet is resolved, we cannot put constraints on the electron density measurement other than to say it is most likely very low, because the line ratio ($F_{\mathrm{\OII}\lambda 3729}/ F_{\mathrm{\OII}\lambda 3726}=1.9\pm0.2$) is outside the range predicted by theoretical calculation (see Section \ref{dense} for our assumptions and references).

\section{Discussion}
\label{discuss}

From the analysis performed on these six LGRB host galaxies with 0.846 $< z < $1.289, it is clear how the VLT/X-Shooter spectrograph can observe emission lines which are generally difficult to detect because either the emission 
lines of the object fall in the gap between optical and NIR ranges and/or for the inadequate sensitivity of earlier spectrographs.
X-Shooter allows the detection of a more complete set of emission lines and this is a fundamental prerogative for the determination of the main features of galaxies even at high redshift.

In particular, with the galaxies analyzed in this work, it was possible to determine the values of metallicity and SFR using different indicators, allowing us to add information about the reliability of these methods also at redshift greater than $1$. In addition, with the spectra, which were particularly rich in emission lines, it was possible to obtain the 
extinction value using both the Balmer lines and the gas electron density from the components ratio of the  \OII\, doublet.
Finally, by using the stellar mass values provided by literature or by us, we also studied the mass-metallicity relation and the sSFR for some of these LGRB host galaxies.

The hosts included in our sample are quite diverse in their properties (e.g. two galaxies hosted a dark GRB), however, given our selection methods (see Section \ref{sample1}), the sample is small and inhomogeneous, preventing us to draw strong conclusions regarding the population of GRB hosts as a whole. For the latter studies of complete samples are imperative \citep[e.g.][]{Hjorth2012,Vergani2014,Kruhler2015,Perley2015a}. We therefore discuss our results within the framework of already known properties of LGRB hosts.

\subsection{SFR and Metallicity of 2 dark LGRB hosts}

The high extinction of the afterglow of GRB\,000210 and the non-detection of the optical afterglow of GRB\,000418 put these two LGRBs in the class of dark bursts (but see \citealt{Schady2014}). In the last years, many works tackled the properties of dark burst host galaxies to test whether they were similar to those of non-dark LGRBs, since the presence of a substantial amount of dust could indicate more evolved hosts than those previously usually found in LGRB host galaxy studies. Dark LGRB host galaxies tend to have on average larger stellar masses and luminosities than non-dark LGRB host galaxy population \citep{Hunt2014, Kruhler2011, Perley2013}. Another topic of interest in those studies is to retrieve information on the dust distribution in high redshift star-forming galaxies. The dust that caused the extinction of the burst could just be concentrated in small clouds around the exploding star or along the LGRB line of sight or be largely present in the entire galaxy. \citet{Covino2013} showed that lines of sight towards dark bursts have higher extinctions than the average extinction measured for non-dark population. The extinction inferred from the Balmer line ratio and/or the optical/NIR SED of the host galaxies of GRB\,000210 and GRB\,000418 (we measured Av for this) is low. Their SFR and metallicity determined from the optical emission lines are not high (though see the different values found for GRB\,000210) and typical of LGRB hosts. On the other hand, sub-mm and radio studies \citep{Michalowski08} showed that they have a large amount of obscured SFR. All this can be explained if the optical and submillimeter emission of the hosts are dominated by different populations of stars, as can be in young hosts experiencing a starburst episode (see \citealt{Michalowski08, Perley2013, Perley2014}). If sub-solar\footnote{For solar metallicity we assume the value of $12 + \log (\mathrm{O/H}) = 8.69$ \citep{Asplund2009}.} metallicities are typically found in general also for the dark-LGRB host galaxies, we could have a further evidence that LGRBs prefer low-metallicity environments. Detailed studies on a larger population of dark burst host galaxies are needed.

\begin{figure}
\includegraphics[width=\linewidth]{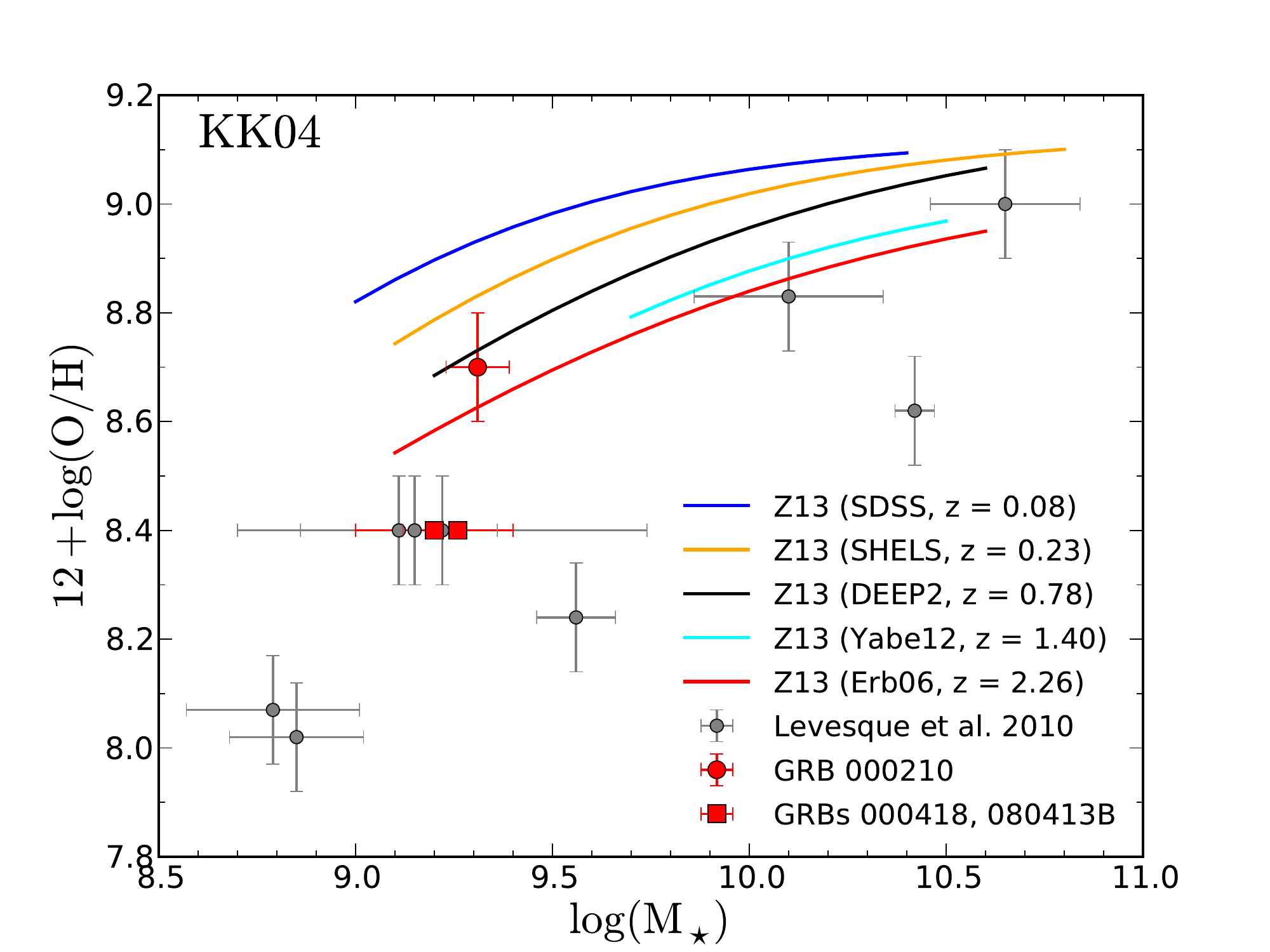}
\caption{Stellar mass-metallicity relation for different galaxy samples. Colored curves are fits to five samples of galaxies lying at different redshifts in the range of $\sim 0.1 - 2.3$ \citep{Zahid13}. Grey circles represent long GRB host galaxies in the redshift interval $0.3 < z < 1$ \citep{levesque10II}. Included are host galaxies of GRB\,000210 (filled point corresponds to the upper-branch solution),  GRB\,000418 and GRB\,080413B, for which we have measurements of both mass and metallicity. All metallicities in this plot correspond to the \citet{Kobulnicky2004} $R_{23}$ diagnostic.}
\label{fig:Mass_Z}
\end{figure}

\begin{figure}
\centering
{\includegraphics[width=\linewidth]{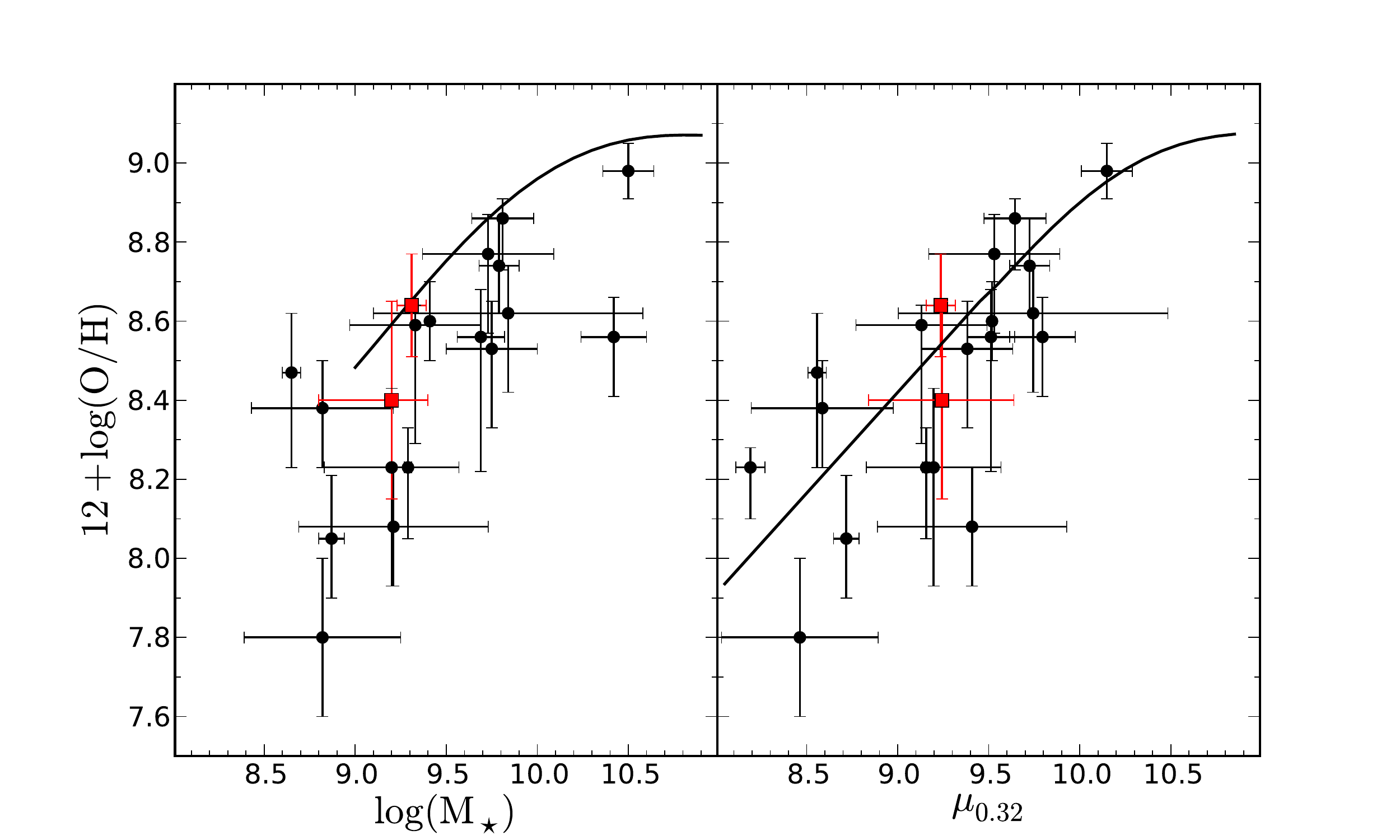}}
\caption{Left: gas metallicity as a function of stellar mass M$_{\star}$. The black points represent the sample of long GRB host galaxies used by \citet{mannucci2011}. The position of GRB\,000210 and GRB\,080413B hosts in the plot is given by the red squares. The black line is a fit to the mass-metallicity relation for local galaxies \citep{mannucci2010}. Right: gas metallicity as a function of $\mu_{0.32} = \log (M_{\star}) - 0.32\log (SFR)$. The black line is the extended fundamental metallicity relation, presented in \citet{mannucci2011}.}
\label{fig:FMR}
\end{figure}

\begin{figure}
\centering
{\includegraphics[width=\linewidth]{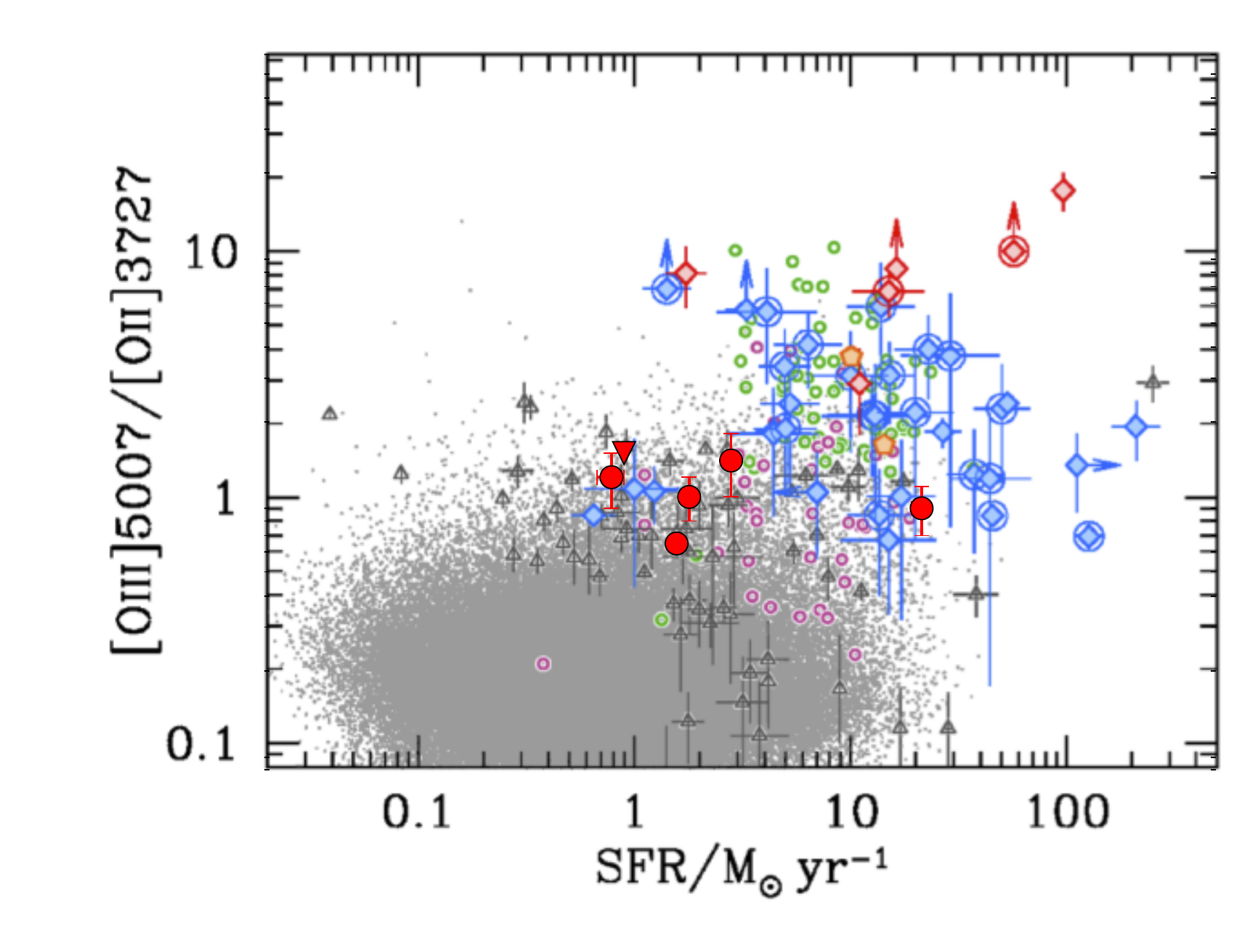}}
\caption{Dependence of \OIIIb$\lambda 5007$/\OII\ on SFR for our six hosts (red points, triangle marks an upper limit for the host of GRB\,000911). Also plotted are data from \citet{nakajima14}: SDSS galaxies (grey points); red and blue diamonds represent $z = 2-3$ Lyman alpha emitters and Lyman-break galaxies, respectively (diamonds with circles denote $z > 3$ galaxies); grey triangles are $z \sim 1$ star-forming galaxies; green and purple circles show Green Pea galaxies and Lyman Break Analogues, while the orange pentagons are Lyman continuum leakers in the Local Universe. To convert our SFR values to \citet{Chabrier2003} IMF, we multiplied them by a factor of 1.2.}
\label{fig:O_ratio}
\end{figure}

\subsection{Mass-Metallicity relation}

In Fig. \ref{fig:Mass_Z} we show the mass-metallicity (M-Z) relation of a number of GRB hosts, including the metallicities of our three LGRB host galaxies for which we measured metallicity and have stellar masses from the literature (GRB\, 000210, GRB\, 000418) and from our measurements (GRB\, 080413B). These values were compared with the results obtained from the analysis that \citet{levesque10II} carried out on a sample of LGRB of redshift z $<$ 1 together with five samples of galaxies lying at different redshifts in the range of $\sim0.1 \sim2.3$ \citep{Zahid13}. All metallicities in this plot correspond to the \citet{Kobulnicky2004} $R_{23}$ diagnostic.

As justified in Section \ref{000210}, in the case of GRB\,000210 we can use the upper branch solution for metallicity determined with the $R_{\rm 23}$ calibration methods. For this host all calibrations return us the near-solar metallicity of $12 + \log (\mathrm{O/H})\sim 8.6$, which is in agreement with the general M-Z relation for star-forming galaxies (see Figure \ref{fig:Mass_Z}). GRB hosts with solar or super-solar metallicity are not unprecedented  \citep{savaglio2010,levesque10II}. In fact, \citet{Kruhler2015}, analysing a large (inhomogeneous) sample of 96 GRB hosts  found that $\sim 20\%$ of GRB hosts have super-solar metallicities. While this demonstrates that not all the environments in which LGRBs explode must necessarily be characterized by low metallicity, a general GRB host population (at least at $z < 1.5$) is still inclined to be metal poor \citep{Graham2013,Vergani2014,Kruhler2015,Perley2015b}. Hosts of GRBs\, 000418 and GRB\, 080413B are found to have metallicities that lie below the general star-forming galaxy population.

\subsection{Fundamental Metallicity Relation}

\citet{mannucci2010} showed that a general relation between stellar mass, metallicity and SFR (i.e., fundamental metallicity relation or FMR) exists at least up to $z \sim 2.5$. The evolution of M-Z relation with redshift, as shown in Fig. \ref{fig:Mass_Z} can therefore be attributed to the fact that galaxies at progressively higher redshifts have higher SFR and consequently lower metallicities. Furthermore, \citet{mannucci2011} showed that the FMR relation also holds for LGRB host galaxies, implying that the stellar mass, metallicity and SFR within these galaxies are related in the same way as in the population of field galaxies. This is not at odds with the findings that LGRBs are preferentially found in galaxies of low metallicities. In Fig. \ref{fig:FMR} we show the FMR for a sample of host galaxies, including the GRB\, 000210 and GRB\, 080413B hosts. The two hosts from our sample are in good agreement with the FMR by \citet{mannucci2011}.

\subsection{Star Formation Rate}

In Fig. \ref{fig:O_ratio} we present the dependence of \OIIIb$\lambda 5007$/\OII\ on SFR for our six hosts (red points, triangle marks an upper limit for the host of GRB\,000911) together with the data from \citet{nakajima14} for several other types of galaxies. The \OIIIb$\lambda 5007$/\OII\ ratio is a proxy for ionization parameter and metallicity: higher oxygen ratio corresponds to lower metallicity and higher ionization parameter (e.g., see Figure 2 in \citealt{nakajima14}). Our six LGRB hosts occupy the same part of the region as a population of star-forming galaxies at $z \sim 1$. However, while other star-forming galaxies are found in a wide parameter space, LGRB hosts are found in a rather narrow region that, according to the photoionization models of \citet{Kewley02}, are consistent with low metallicities and low ionization parameters. 

\section{Summary and conclusions}
\label{summary}
We present VLT/X-Shooter data and results of 6 GRB host galaxies in the redshift range of $0.8 < z < 1.3$. Thanks to X-Shooter capabilities and resolution, we were able 
to measure the common bright emission lines in the majority of our hosts in order to better characterize the 6 GRB host galaxies at $z < 1$. From the emission line measurements we were able to derive properties like dust extinction, SFR, metallicities and electron densities where possible. 
In particular we measured SFR and metallicities for all the hosts of the sample. Only three host masses where available, therefore, regarding the M-Z relation based on the three hosts measurements, we found that one host fall in the general M-Z relation of star-forming galaxies and two of them are found to lie below this relation. These hosts
have metallicities around solar or below solar, which is consistent with the growing evidence for a metallicity threshold around this value. The two dark GRB hosts are found to have low host-averaged extinctions and star formation rates, derived from optical emission lines, that are much lower than the ones measured from the sub-mm and radio studies \citep{Michalowski08}. This may be an indication of different stellar populations and ISM properties which contributed to the emission at different wavelengths and shows the necessity for multiwavelength studies if we are to understand the hosts' properties in detail. Moreover, the two hosts for which we have stellar mass, metallicity and SFR are found to be in good agreement with the FMR relation by \citet{mannucci2011}. Comparing SFR measurements from our 6 hosts with several samples of galaxies we found that they occupy the same region as a star-forming galaxy population at $z \sim 1$.

\section*{Acknowledgments}
This work is based on VLT/X-Shooter spectroscopy observations performed at the European Southern Observatory, Paranal,
Chile. We acknowledge ESO personnel for their assistance during the observing runs and Roberto Pallavicini, the Italian P.I. of X-Shooter, who was one of the most active promoters of this wonderful instrument.
We thank Tommaso Vinci with his software Identifica.
SP acknowledges partial support from PRIN MIUR 2009 and from ASI INAF I/004/11/1.
E.Pian acknowledges financial support from MIUR and Scuola Normale Superiore, and from grants ASI INAF I/088/06/0 and INAF PRIN 2011.
This research has made use of the GHostS database (www.grbhosts.org), which is partly funded by Spitzer/NASA grant RSA Agreement No. 1287913.

\bibliographystyle{aa}
\bibliography{paper_references}

\label{lastpage}
\end{document}